\documentclass{kapproc} % Computer Modern font calls
\usepackage{graphicx}

\let\footnote\savefootnote

\let\footnotetext\savefootnotetext 

\kluwerbib
\begin{document}

%------------ article title  ------------------->>

% If you use \\'s , please supply an alternate version of the title
% in square brackets, i.e., 
%\articletitle[Communism, Sparta, and Plato]
%{COMMUNISM, SPARTA,\\ and PLATO}

\articletitle[The VLBA 2cm Survey]{THE VLBA \protect{2cm} SURVEY:\\ 
KINEMATICS OF \protect{pc}-SCALE STRUCTURES IN ACTIVE GALACTIC 
NUCLEI
}

%% optional, to supply a shorter version of the title for the running head:
%%\chaptitlerunninghead{}

%\author{
%E. Ros$^1$,
%K.I. Kellermann$^2$,
%M.L. Lister$^2$,
%J.A. Zensus$^1$,
%M.H.\ Cohen$^3$,
%R.C. Vermeulen$^4$
%M. Kadler$^1$,
%D.C.  Homan$^2$
%}
%\affil{{\rm 1}: Max-Planck-Institut f\"ur Radioastronomie, Bonn, Germany\\
%{\rm 2}: National Radio Astronomy Observatory, Charlottesville, VA, USA\\
%{\rm 3}: ASTRON, Dwingeloo, The Netherlands\\
%{\rm 4}: California Institute of Technology, Pasadena, CA, USA}

%, Auf dem H\"ugel 69, D-53121 Bonn, Germany
%, 520 Edgemont Road, Charlottesville, VA 22903, USA
%Netherlands Foundation for Research in Astronomy, Postbus 2, NL-7990 AA
%Dwingeloo, The Netherlands
%Department of Astronomy, MS 105-24, 
%, Pasadena, CA 91125

\author{E. Ros}
\affil{
Max-Planck-Institut f\"ur Radioastronomie, 
Auf dem H\"ugel 69, D-53121 Bonn, Germany
}
\email{ros@mpifr-bonn.mpg.de}

%\author{K. I. Kellermann, M. L. Lister, D. C. Homan}
%\affil{
%NRAO,
%520 Edgemont Road, Charlottesville, VA 22903, USA
%}
%
%\author{M. H. Cohen}
%%\affil{
%Dept.\ of Astronomy, MS 105-24, 
%Caltech, Pasadena, CA 91125, USA
%}
%
%%\author{J. A. Zensus, M. Kadler}
%%\affil{
%%Max-Planck-Institut f\"ur Radioastronomie, 
%%Auf dem H\"ugel 69, D-53121 Bonn, Germany
%%}
%
%
%\author{R. C. Vermeulen}
%\affil{
%ASTRON, Postbus 2, NL-7990 AA
%Dwingeloo, The Netherlands
%}
%
%%% multiple authors may be separated with \\
%%% \author{Samuel Bostaph\\
%%% and Gregor Kariotis}

% optional prologue
%\prologue{<text>}{<author, year>}

% optional abstract
 \begin{abstract}
The kinematics of jets in active galactic nuclei (QSOs, BL\,Lacs,
Radio Galaxies and Empty Field objects) on parsec scales is being
studied with Very Long Baseline Array (VLBA) observations at 15\,GHz of
a sample of more than 170 radio sources.
More than 1000 images have been taken since 1994.  Here we present
an overview of the results of our study, including the proper
motions of components in the jets, and their relationship with
other source properties.
 \end{abstract}

% optional keywords
% \begin{keywords}
% Text, text...
% \end{keywords}

%------------ body of article ------------------->>
\paragraph{Introduction}
The emission at radio frequencies from active galactic nuclei (AGN) usually 
presents the form of collimated ``jets" which connect a compact central
region (the ``engine" of the AGN) with kiloparsec-scale extended lobes
and hot spots.  The technique of Very Long Baseline Interferometry (VLBI)
is the standard tool to study the compact structure of these jets at
parsec scales, which corresponds to resolutions of the order of 
1 milliarcsecond.  
Relativistic motion oriented nearly along the line of sight
causes an aparent superluminal motion due to the compression
of the time frame.
There have been different surveys to study the nature of AGN
physics.  Those studies include VLBI observations at cm and mm-wavelengths,
single-dish radio flux monitoring, and observations in the optical, X-ray
and $\gamma$-ray.
To complement those,
we have embarked on the project of monitoring a sample of radio
sources with the VLBA at a 2\,cm wavelength, which offers a good compromise
between the detection of the extended jet features (fainter at higher 
frequencies because of their optically thin spectrum) and better insight
close to the jet ``core" (opacity$\sim$1 region) 
at a relatively high frequency.

\paragraph{The Survey}
Our program consists in monitoring more than 170 radio sources using
the VLBA at 15\,GHz.  We have been observing since
late 1994, with more than 40 VLBA runs.  With typically 24 images per
run, we have over 1000 images to date. 
The final goal of our research is the characterization
of the kinematics of AGN jets and their relationship to other source properties.
The definition of the sample and contour maps of each
source are given in
Kellermann et al.\ (1998) and Zensus et al.\ (2002).  Overall,
60\% of the observed
AGN are QSOs, a 20\% are BL\,Lac Objects, and less 
than a 15\% are Radio Galaxies.

Each image of the survey is derived from an observation covering 8\,hr
of hour angle with a total of 50\,min integration time per source
at each epoch.
The root-mean-square
noise level in the images is less than 1\,mJy\,beam$^{-1}$.  
Consequently, the dynamic ranges reached are 1:1000, which is enough
to measure the kinematics of the objects.  The synthesized beam of the
VLBA at 15\,GHz is typically of 0.5\,milliarcseconds (mas)
in size.  Automatic imaging was carried out in general (applying loops
of {\sc clean} and phase self-calibration) using the {\sc difmap}
software.

An  online archive with the results of the survey is available under
{\tt http://www.nrao.edu/2cmsurvey}.

We measured the positions of the absolute and relative peaks of 
brightness using the task {\sc jmfit} in $\cal{AIPS}$.
We cross-identified the components between different epochs and with
this database we have determined the component motions for 
more than 100 sources in 
our sample extending over a time baseline of 4 to 6 years.
We have studied particular sources in more detail, such as
NGC\,1052 (0238$-$084; Vermeulen et al.\ 2002)
or 4C\,+12.50 (1345+125; Lister et al.\ 2002).

\begin{figure}[htb!]
\begin{center}
\includegraphics[angle=-90,width=1.0\textwidth]{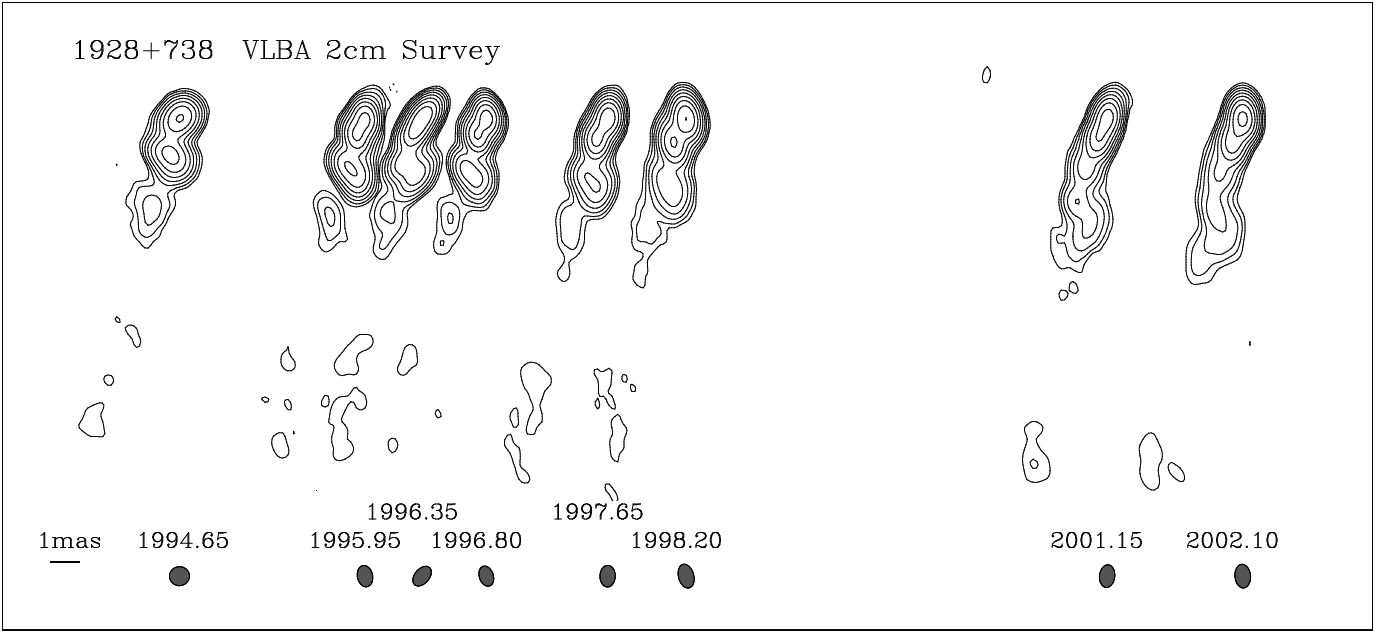}
~\\[3mm]
%\vspace*{20pt}
\includegraphics[clip,width=1.0\textwidth]{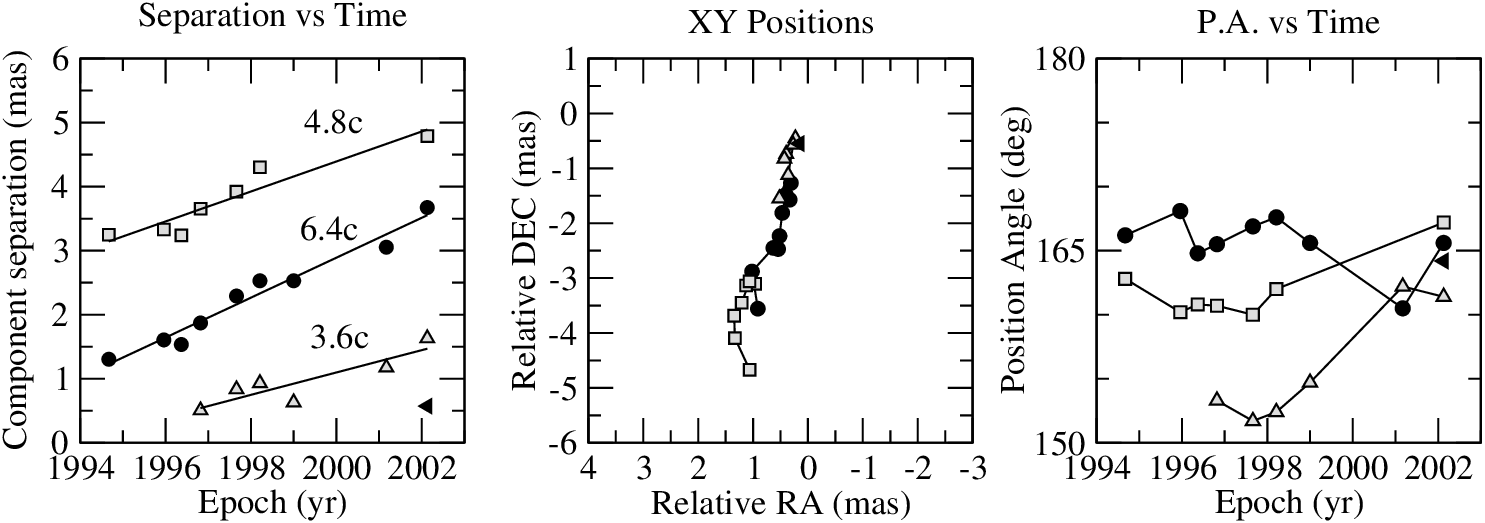}
\vspace*{-20pt}
\end{center}
\caption[]{Top panel: Contour images from the QSO\,1928+738 from the survey
database.
%, aligned to the peak of brightness for the eight epochs
%observed to date.  
The interferometric beams are below the
epoch label.  The contours are 
$6\times(1,2,4,8,...)$\,mJy\,beam$^{-1}$.
Bottom, left panel: Component separation from the brightness peak 
against time.  The proper motions are of 0.17, 0.31, and 
0.23\,mas\,yr$^{-1}$ from the inner to the outer components,
respectively, which correspond
to the plotted apparent speeds.
($\Omega_m=0.3$, 
$H_0=65$\,km\,s$^{-1}$\,Mpc$^{-1}$ is used here and in Fig.~2).
Bottom, middle panel: Motions in the plane of the sky of the measured 
components.
Bottom, right panel: Position angle against time.
%Bottom, right panel: Comparison of the flux density evolution measured at
%the University of Michigan monitoring program (UMRAO, supported by 
%funds from the University of Michigan)
%and obtained from the
%VLBA imaging (sum of the {\sc clean} components).
\label{fig:ros1_1928}
}
\end{figure}

\paragraph{An Example: 1928+738}
As an example of our analysis, we show 
in Fig.~\ref{fig:ros1_1928} the results on the QSO\,1928+738 (4C\,73.18)
from our image database, based on 8 epochs of observations.
The images in the top panel have been
aligned relative to the brightness peak, although it is known from
astrometric results (see Ros et al.\ 1999 and references therein)
that the central engine is somewhere to the north of the brightest component
at frequencies lower than 15\,GHz.
The bottom panels show the estimated positions of
the relative peaks of brightness in the images with respect to the
main one (associated tentatively with the core). 
The kinematical analysis based on the procedure described above
shows that the components move southwards 
with superluminal speeds in slightly
curved trajectories.  
%Those results confirm the rate of emergence
%of components of one every 1.6\,years reported by Hummel et al.\ (1992)
%and the space-VLBI results at 5\,GHz.

\begin{figure}[htb!]
\begin{center}
\includegraphics[clip,width=0.81\textwidth]{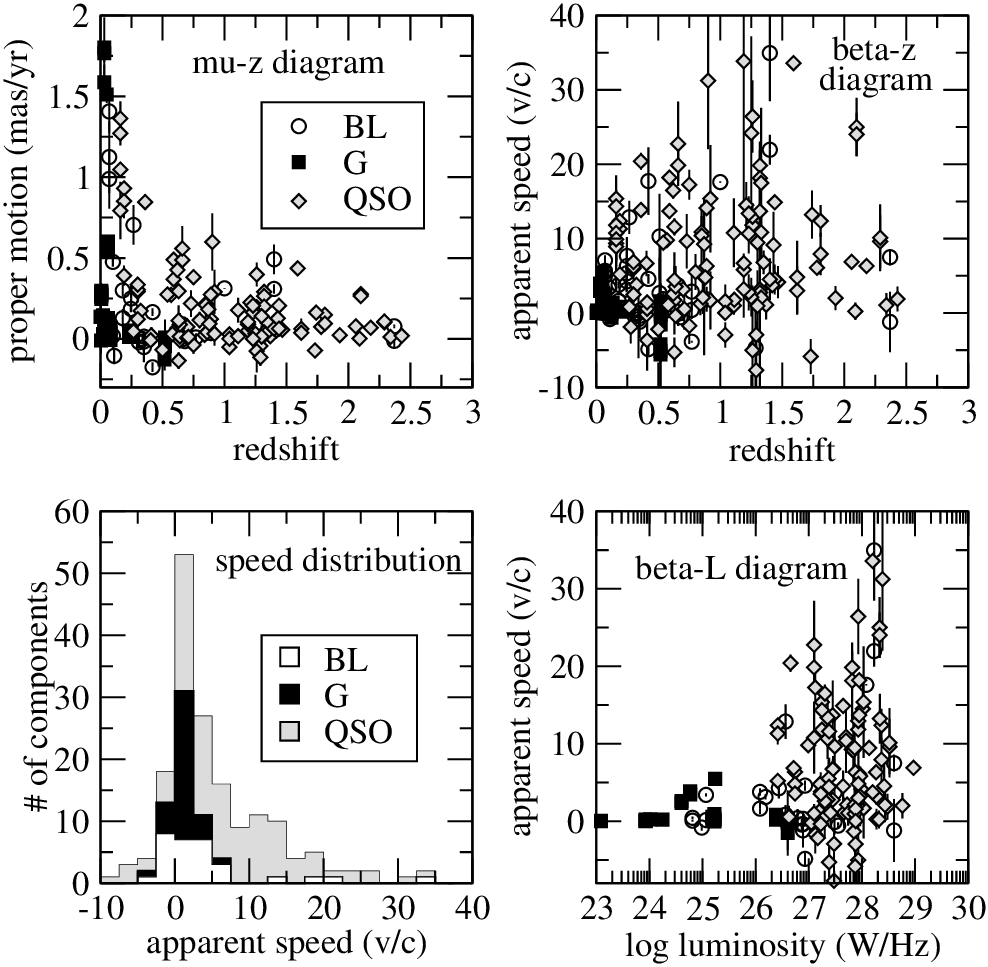}
\vspace*{-20pt}
\end{center}
\caption{Preliminary overall values of the statistical analysis of our
survey.  
%Left panel: Distribution of measured flux densities with redshift.
Top, left panel: $\mu$-z diagram, proper motions against redshift for the
best measured components in the survey (30 components for BL\,Lac objects,
35 for radio galaxies and 105 for QSOs).  
Top, right panel: Apparent speeds against redshift for the same
components.  Bottom, left panel: Distribution of the
deduced apparent speeds from the proper motions.
Bottom, right panel: Apparent speeds against total source radio luminosity
at 15\,GHz.
%($\Omega_m=0.3$, $H_0=65$\,km\,s$^{-1}$\,Mpc$^{-1}$ has been used).
\label{fig:ros1_stat}
}
\end{figure}

%\nopagebreak[4]
\paragraph{Statistics and kinematics}
A detailed description of the kinematic analysis on the 
radio sources will be given in Kellermann et al.\
(in preparation), but we give some preliminary conclusions here.
%The left panel of 
%Fig.~\ref{fig:ros1_stat} shows the luminosity versus redshift
%for each source with measured speeds
%(one epoch per source, flux density obtained
%as the sum of the {\sc clean} components).
In general, radio galaxies have 
subluminal proper motions, and QSOs have faster motions in mean than
BL\,Lac objects.  Notice also that the BL\,Lac objects have generally
lower redshifts.  From the bottom, right panel, a correlation between
luminosity and apparent jet speed is evident.  
Especially remarkable is the lack of high velocity jets with low luminosities.
See Lister et al.\
(2003) for a discussion on this subject.

\paragraph{MOJAVE}
We are extending now our program in a long-term project named MOJAVE 
(MOnitoring of Jets in A{\sc gn} with V{\sc lba} Experiments).  
The observations were begun in August 2002.
This observational effort includes also linear and circular polarization
studies of the AGN.  
%A revision of the sample membership, seeking
%completeness based on flux density, is currently underway.

%% optional
%\begin{acknowledgments}
\begin{small}
\paragraph{\small Acknowledgments}
This paper is based on work done in collaboration with
M.H.\ Cohen,
D.C.\ Homan, 
M.L.\ Lister, 
M.\ Kadler,
K.I.\ Kellermann, 
R.C.\ Vermeulen,
and
J.A.\ Zensus. 
The Very Long Baseline Array of 
the USA National Radio Astronomy Observatory is operated by 
Associated Universities, Inc., under cooperative agreement with the
USA National Science Foundation.
\end{small}
%\end{acknowledgments}

%or 

\end{document}